\documentclass[a4paper]{jpconf}

\usepackage{iopams}

\usepackage{graphicx}

\newcommand{\hp}{(C$_5$H$_{12}$N)$_2$CuBr$_4$}

\begin{document}

\title{Thermal expansion of the spin-$\frac{1}{2}$ Heisenberg-chain compound Cu(C$_4$H$_4$N$_2$)(NO$_3$)$_2$}

\author{J Rohrkamp$^1$, M D Phillips$^2$, M M Turnbull$^2$, and T Lorenz$^1$}

\address{$^1$ II. Physikalisches Institut, Universit\"at zu K\"oln, Germany}
\address{$^2$ Carlson School of Chemistry and Biochemistry, Clark University, USA}

\ead{rohrkamp@ph2.uni-koeln.de}

\begin{abstract}
Compounds containing magnetic subsystems representing simple model
spin systems with weak magnetic coupling constants are ideal
candidates to test theoretical predictions for the generic
behavior close to quantum phase transitions. We present
measurements of the thermal expansion and magnetostriction of the
spin-$\frac{1}{2}$-chain compound copper pyrazine dinitrate
Cu(C$_4$H$_4$N$_2$)(NO$_3$)$_2$. Of particular interest is the
low-temperature thermal expansion close to the saturation field
$H_c \simeq 13.9\,\mathrm{T}$, which defines a quantum phase
transition from the gapless Luttinger liquid state to the fully
saturated state with a finite excitation gap. We observe a sign
change of the thermal expansion for the different ground states,
and at the quantum critical point $H_c$ the low-temperature
expansion approaches a $1/\sqrt{T}$ divergence. Thus, our data agree very well with the expected quantum critical behaviour.
\end{abstract}

%\section{Introduction}

Copper pyrazine dinitrate Cu(C$_4$H$_4$N$_2$)(NO$_3$)$_2$ (or
CuPzN) crystallizes in an orthorhombic structure with lattice
constants $a=6.712\,\mathrm{\AA}$, $b=5.142\,\mathrm{\AA}$ and
$c=11.732\,\mathrm{\AA}$, space group $Pmna$. The structure consists
of linear Cu-pyrazine-Cu-chains along the $a$
axis. \cite{Santoro70} By zero-field susceptibility and specific
heat measurements it could be shown, that CuPzN is a $S=1/2$ chain
with an antiferromagnetic (AFM) intrachain exchange constant of
$J/k_\mathrm{B}\simeq 10.6\,\mathrm{K}$. \cite{Hammar99} Recent
zero-field muon-spin relaxation measurements provide evidence for
long-range magnetic order below $T_\mathrm{N} \simeq
100\,\mathrm{mK}$, which implies a ratio of interchain to
intrachain coupling of $J'/J \simeq 4.4\cdot
10^{-3}$. \cite{Lancaster06} This means, that the spin chains are
well isolated from each other and CuPzN very well realizes the
model of a one-dimensional spin-1/2 Heisenberg chain
antiferromagnet. This theoretical model is of particular interest
because it represents a so-called spin Luttinger liquid (LL) with
a continuous two-spinon excitation spectrum. For a finite
magnetoelastic coupling, however, any AFM spin-1/2 chain is
expected to show a Spin-Peierls transition, which transforms the
LL to a dimerized phase with a finite excitation gap. Apparently,
the magnetoelastic coupling in CuPzN is low enough to prevent such
a Spin-Peierls transition for $T>T_\mathrm{N}$. Because
$k_\mathrm{B}T_\mathrm{N} \ll J$, CuPzN is ideally suited for
experimental studies of the LL state over a wide range of
temperature and due to the relatively low value of $J$ it also
allows one to  study the magnetic-field influence over a wide field
range. The zero-temperature quantum phase transition from the LL
phase to the fully spin-polarized high-field phase with a finite
spin gap is expected to occur at $H=2J/g\mu_\mathrm{B}\sim
15.8$~T, which is accessible in typical superconducting laboratory
magnets. On approaching this quantum critical point, highly
anomalous temperature dependencies of various thermodynamic
properties are expected~\cite{Zhu03,GarstRoschPRB} and have been
observed recently in the related spin-1/2-ladder compound
\hp.~\cite{Lorenz08,Anfuso08,RueggPRL08} In this report, we
present high-resolution measurements of the thermal expansion
$\alpha(T,H)$ and the magnetostriction in magnetic fields up to
17~T. For both zero field, as well as the quantum critical field,
our data well agree with the theoretical expectations, namely
$\alpha(T,H=0)$ scales with the calculated magnetic specific heat
while for the critical field it shows a power-law divergence
$\alpha(T,H=H_c)\propto -1/\sqrt{T}$.

%\section{Experiment}

\begin{figure}[t]
\begin{center}
\includegraphics[width=.43\textwidth]{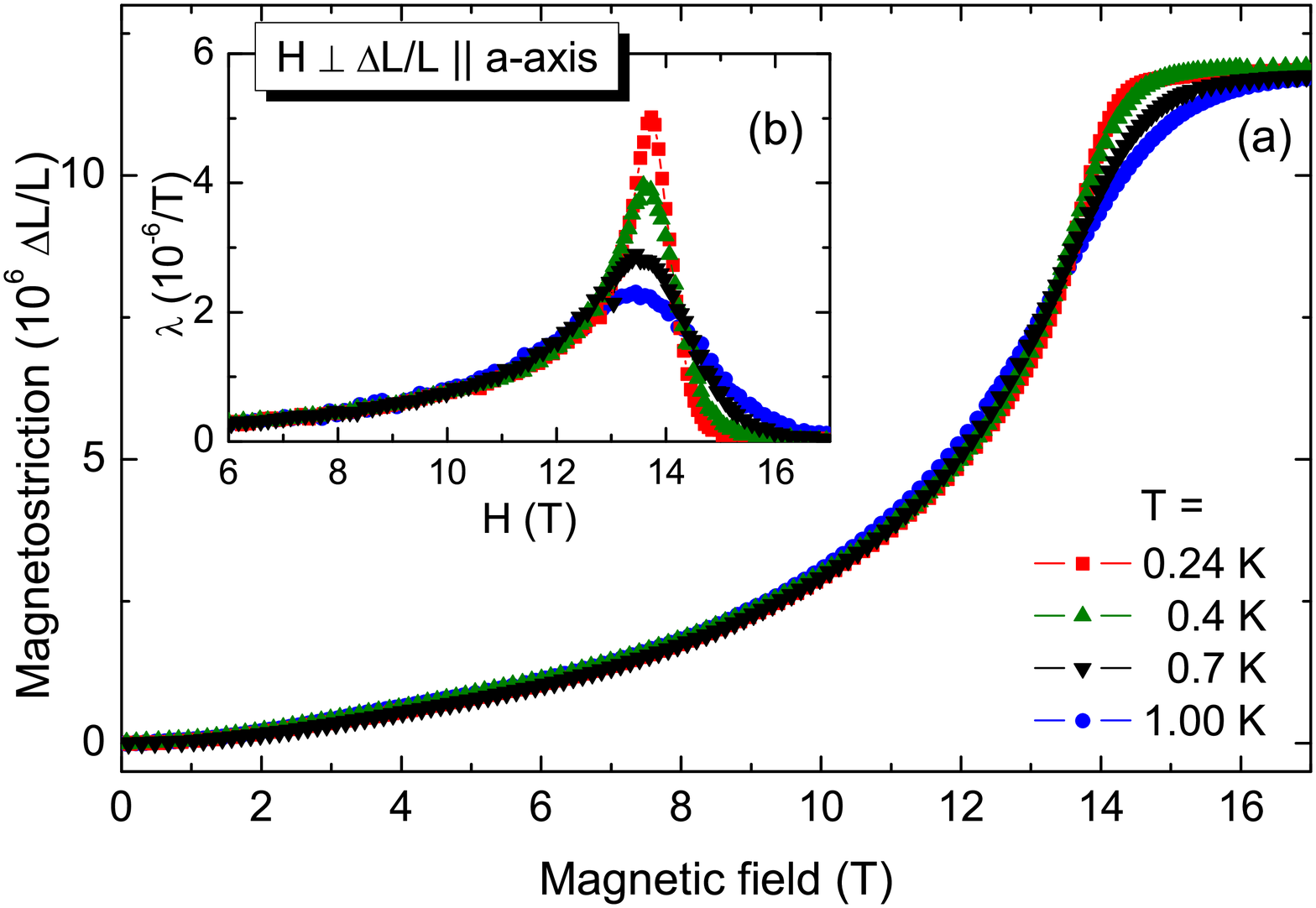}
%\end{center}
\hspace{1pc}
%\begin{minipage}[b]{14pc}
\includegraphics[width=.42\textwidth]{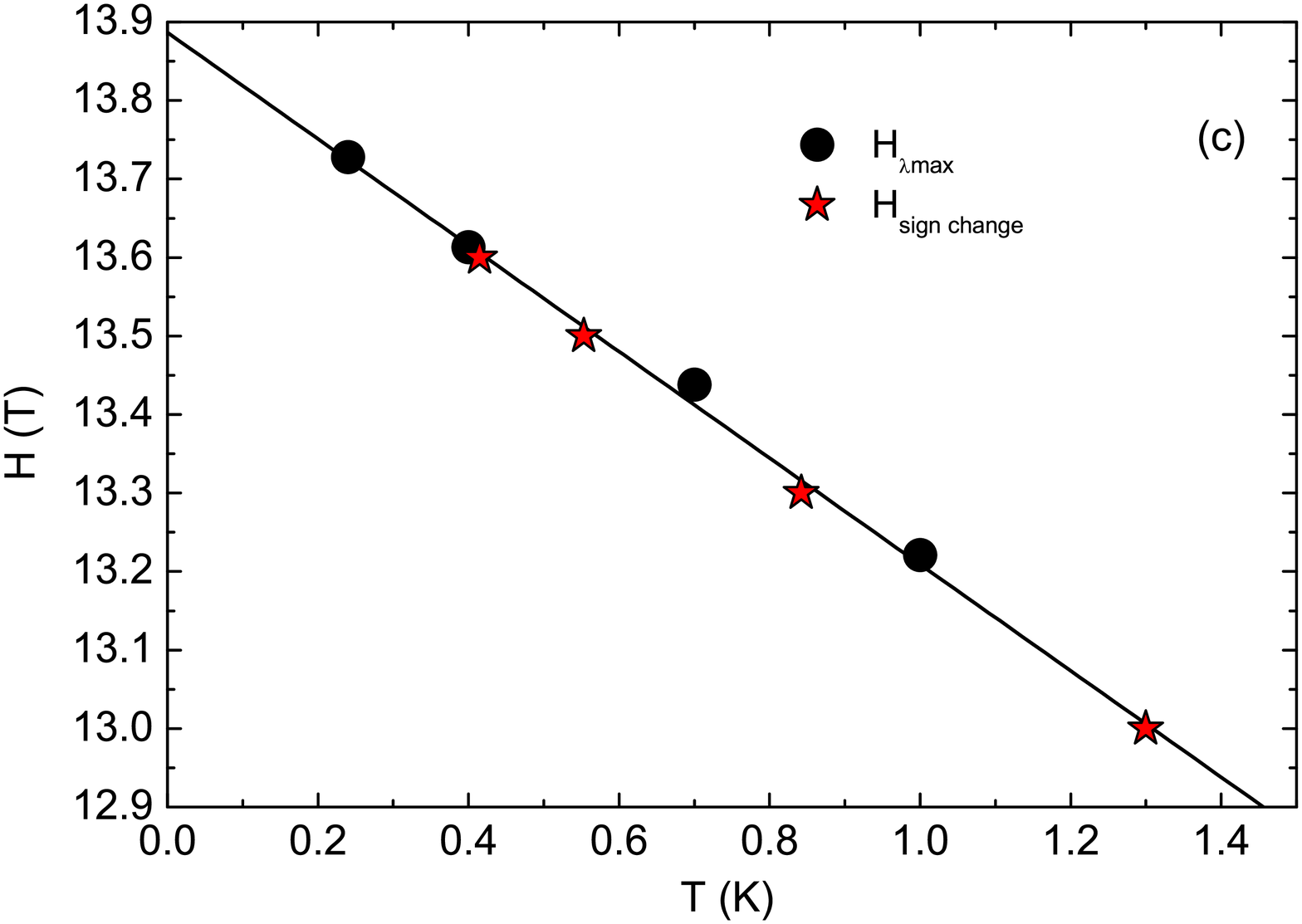}
\end{center}
%\caption{\label{MSa}Magnetostriction $\varepsilon = \Delta L/L$
%along a-axis at various constant temperatures with magnetic field
%along b-axis. Inset: $\lambda = \partial\varepsilon / \partial H$}
%\end{minipage}
\caption{\label{MSa}(a) Magnetostriction $\varepsilon = \Delta
L(H)/L_0$ and (b) its field derivative
$\lambda=\partial\varepsilon / \partial H$ measured along the spin
chains. (c) Determination of $H_c^{\| b}(T=0)$ via a linear
extrapolation of the peak positions ($\bullet$) of $\lambda$ and the
points ($\star$) where $\alpha(T,H)$ changes sign (see text).}
\end{figure}

The magnetostriction and thermal expansion was measured in a
home-built capacitance dilatometer,
%The sample is mounted on a screw-platform in contact
%with the moveable plate of a capacitor. Thus, the length change
%$\Delta L$ of the sample is transformed into the distance change
%of the capacitor plates and determined by the change of the capacity. The dilatometer
which is attached to a $^{3}\mathrm{He}$ insert ($T_{\rm min}\simeq
250$~mK) of a $^{4}\mathrm{He}$ bath cryostat equipped with a
superconducting magnet ($H_{\rm max}= 17$~T). The dilatometer can
be rotated with respect to the magnetic-field axis such that the
field can be aligned in any angle with respect to the measured
uniaxial length change $\varepsilon(T,H)=\Delta L(T,H)/L_0$. For
magnetostriction (thermal expansion) measurements $T$ ($H$) is
kept fixed while $H$ ($T$) is continuously varied. $\Delta L(T,H)$
is measured with respect to the length at $H=0$ ($T=T_{\rm min}$)
and $L_0$ is the total length of the sample. The derivatives
$\lambda=\partial\varepsilon / \partial H$ and
$\alpha=\partial\varepsilon / \partial T$ are obtained
numerically. The sample studied here is from the same batch as
those used in Ref.~\cite{Sologubenko07}. We measured the length
changes parallel to the spin chains, i.e.\ parallel to the $a$
axis, on a crystal with $L_0=2.6$~mm and the perpendicular
dimensions $b\times c \simeq 0.4 \times 0.7$~mm$^2$. The magnetic
field was oriented along the $b$ axis, which has the largest $g$
factor ($g_b=2.27$~\cite{Hammar99}) in order to minimize the
critical field $H_c^{\| b}\simeq 13.9$~T.

%\section{Results and Discussion}

Figure\,\ref{MSa}(a) shows the magnetostriction at various
temperatures as a function of field, which strongly resembles the
behaviour of the low-temperature magnetization \cite{Hammar99}.
With increasing field the sample continuously elongates and
finally reaches a plateau above the saturation field.
Qualitatively, this magnetostriction can be explained as follows: as the
magnetic field forces the spins to orient parallel, the lattice
distorts in a way to decrease the AFM coupling $J$.
This reduces the cost in exchange energy, but causes an
additional cost in elastic energy. Because, in lowest order, the
first (second) term is linear (quadratic) in the distortion, such
a magnetostriction always occurs. In the present case, $J$ is
minimized by increasing the distance between neighboring spins,
which appears rather natural, but there are also cases where the
magnetostriction has the opposite sign~\cite{Johannsen05}.
The uniaxial
pressure dependence of $J$ is related to the saturation value of
the magnetostriction $\varepsilon_s(H>H_c)\simeq 1.2\cdot 10^{-5}$
via $\partial J/\partial p_a = \mathcal{D}^{-1}\cdot \varepsilon_s\,V_{\rm fu}$. Here,
$p_a$ means pressure along $a$, $\mathcal{D}\simeq 0.69$ is the field-induced change of the spin corellator and $V_{\rm fu}\simeq 203$~\AA$^3$
is the volume per formula unit.~\cite{Anfuso08,Remark} This yields $\partial \ln
J/\partial p_a \simeq 2.4$~\%/GPa.

As shown in Figure\,\ref{MSa}(b) the field derivative $\lambda $
has a rather sharp peak, which systematically broadens and shifts
towards lower field with increasing temperature. The peak
positions are plotted as a function of temperature in
Figure\,\ref{MSa}(c) and a linear fit extrapolates to the
zero-temperature critical field $H_c^{\| b}\simeq 13.9$~T. This
method to determine the QCP has already been used in the
spin-ladder system \hp.~\cite{Anfuso08}

\begin{figure}[t]
\begin{center}
\includegraphics[width=.36\textwidth]{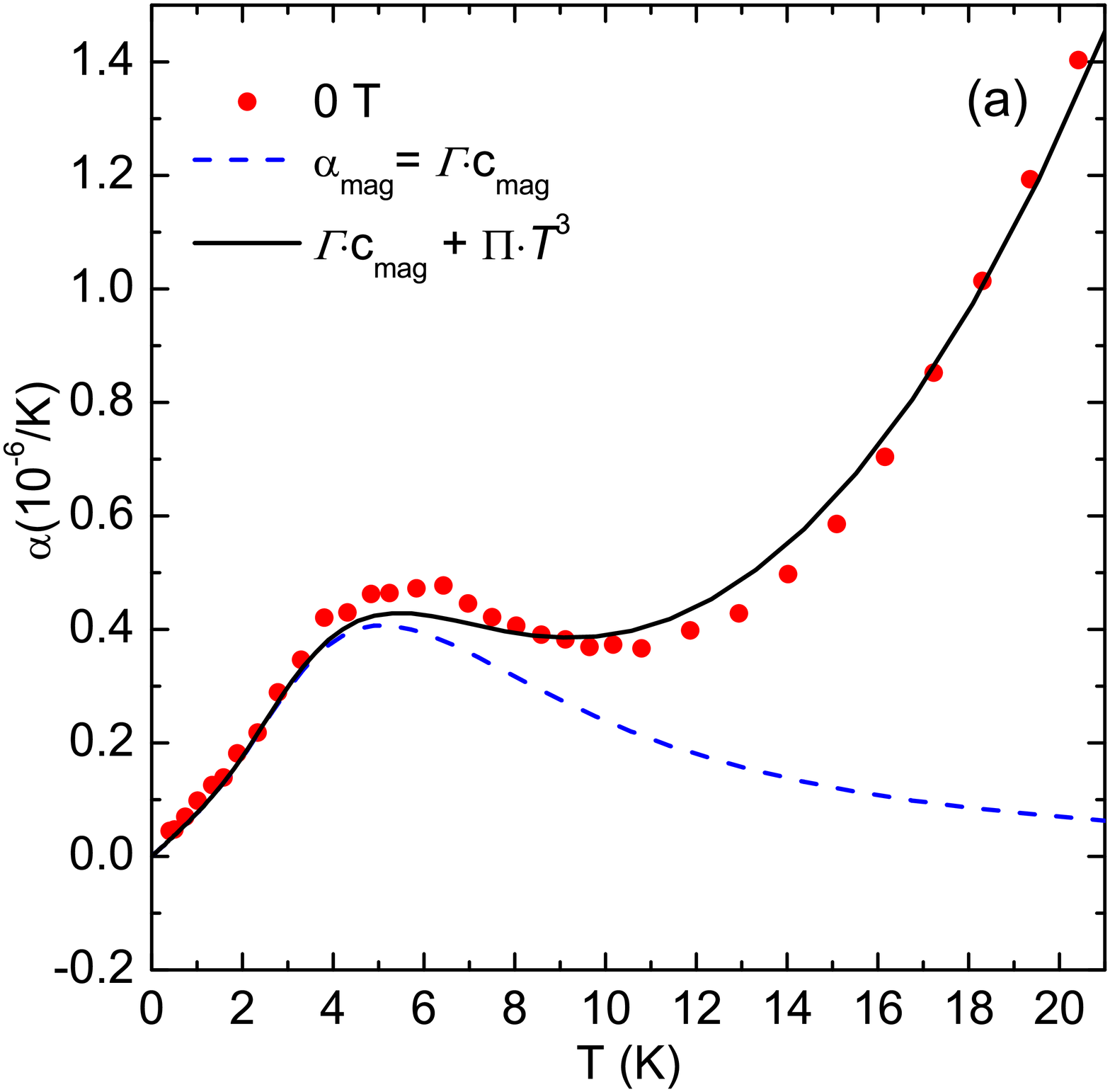}\hspace{1pc}
\includegraphics[width=.36\textwidth]{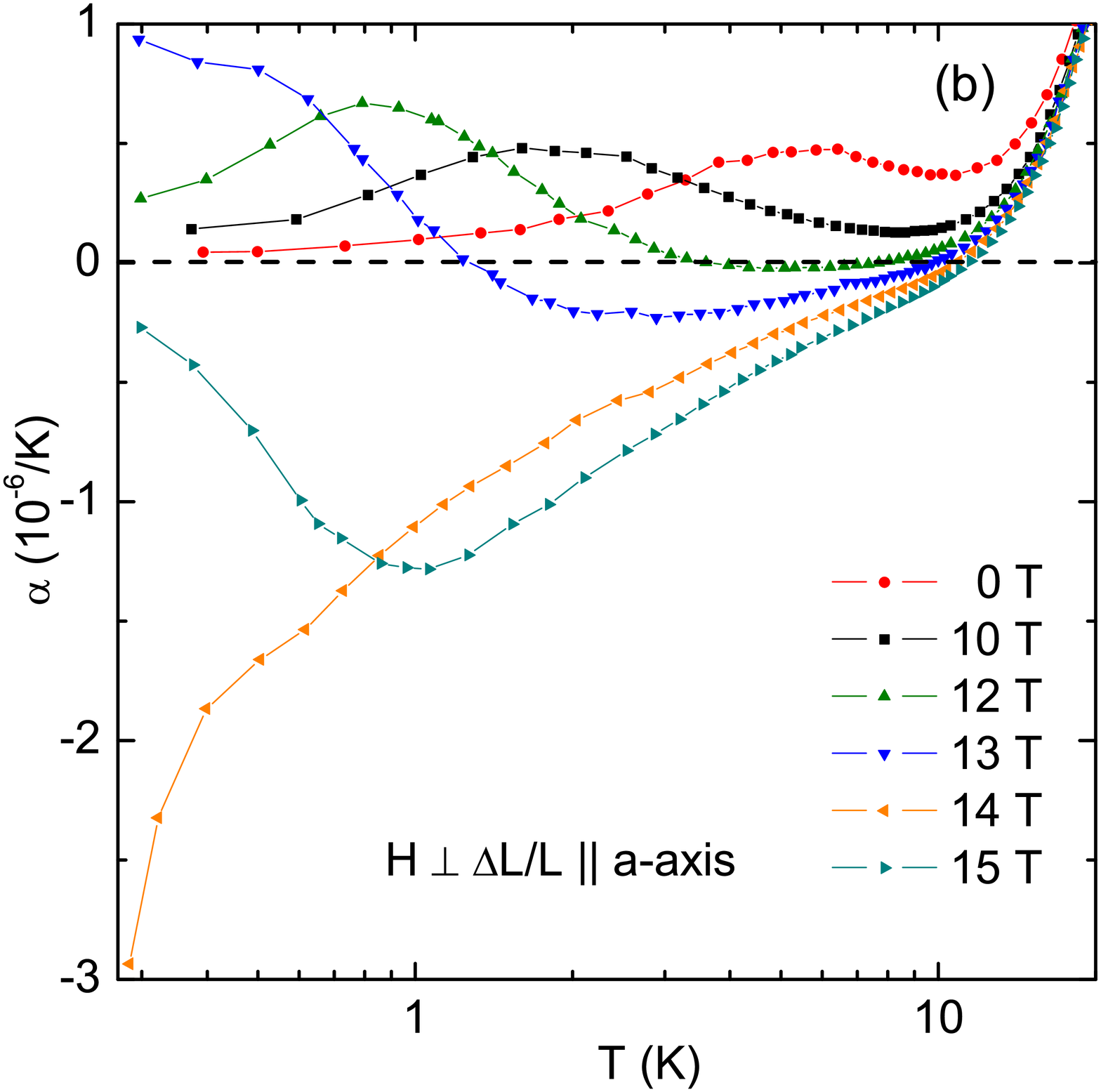}
\end{center}
\caption{\label{TAD} Thermal expansion along the spin chains
measured in $H=0$ and $H\le 15$~T; note the logarithmic $T$ scale
in (b). The solid line in (a) is the sum of the scaled magnetic
specific heat of a Heisenberg chain (dashed) and a phononic $T^3$
term (see text).}
\end{figure}

Figure\,\ref{TAD}(a) displays the zero-field $\alpha (T)$
showing a maximum around 6\,K and a strong increase above
11\,K. The latter is the phononic thermal expansion ($\alpha_{\rm
ph}$), while the maximum is of magnetic origin ($\alpha_{\rm
mag}$). As the Heisenberg chain contains only one energy scale
$J$, a Gr{\"u}neisen scaling between the magnetic
contribution $c_{\rm mag}$ of the specific heat and $\alpha_{\rm
mag}$ is expected, i.e.\ $\alpha_{\rm mag}(T)=\Gamma \cdot c_{\rm
mag} (T)$~\cite{Zhu03,Lorenz07JMMM}.
%From the magnetostriction data
%$\Gamma= \partial \ln J/\partial p_a \simeq 1.7$~\%/GPa is known
%and we assume the typical $T^3$ dependence for the low-temperature
%expansion.
Using the result of $c_{\rm mag}(T)$ for the
Heisenberg chain~\cite{Kluemper00}, and assuming the usual $\alpha_{\rm
ph}\propto T^3$ of acoustic phonons, we can model the measured
thermal expansion $\alpha=\Gamma c_{\rm mag} (T) +\Pi\, T^3$ by
adjusting $\Gamma$ and $\Pi$ as free parameters. With $\Gamma= \partial \ln J/\partial p_a = 1.7$~\%/GPa and $\Pi=1.5\cdot 10^{-10}/\mathrm{K}^4$, we
obtain the solid line in Figure\,\ref{TAD}(a), which nicely describes
the experimental data up to 20~K and the corresponding
$\alpha_{\rm mag}$ is given by the dashed line. Obviously, the
low-temperature expansion $\alpha (T<4$~K) of CuPzN is of almost purely
magnetic origin. The fact that this pressure dependence amounts only to about 2/3 of the value obtained from the magnetostriction data, is most likely due to the uncertainty of the phononic background, but does not affect the following discussion~\cite{Remark}.

Figure\,\ref{TAD}(b) gives an overview of the thermal expansion
data up to 15~T. With increasing field the maximum of $\alpha$ shifts
to lower temperature and above 10~T its amplitude increases on
further approaching the critical field $H_c^{\| b}\simeq 13.9$~T.
For a slightly higher field of 14~T, however, the thermal
expansion is completely different; $\alpha(T)$ is negative in the
entire low-$T$ range and monotonically decreases down to the
lowest temperature. For an even higher field of 15~T, $\alpha(T)$
finally shows a minimum around 1~K.

\begin{figure}[t]
\begin{center}
\includegraphics[width=.89\textwidth]{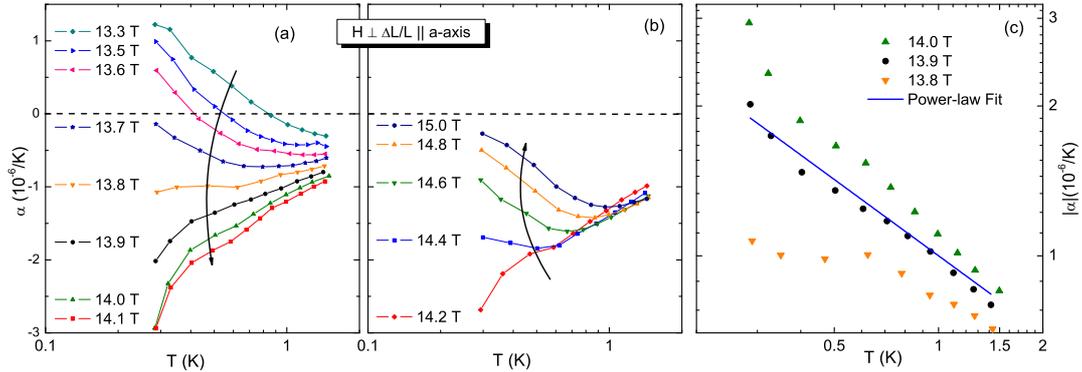}
\end{center}
\caption{\label{TADTT} Thermal expansion along the spin chains for
fields close to $H_\mathrm{c} \approx 13.9\,\mathrm{T}$. The line
in (c) is a power-law fit yielding $\alpha(T,13.9~{\rm T})\propto
T^{-0.51}$ (see text).}
\end{figure}

Figure\,\ref{TADTT} gives a detailed view on the field region
around the QCP. As shown in panel~(a), the $\alpha(T)$ curves
systematically change their curvature from an upward to a downward
bending when the magnetic field is increased from 13.3~T to 14.1~T.
From this figure alone, the location of the QCP is not obvious.
Note, however, that only the $\alpha(T)$ curves for $10~\mathrm{T}< H
<13.7$~T show a sign change as a function of $T$; see also
Figure\,\ref{TAD}(b). Interestingly, the points where
$\alpha(T^\star,H^\star)=0$ match the line obtained from the
maximum positions of the $\lambda(T,H)$ curves, see
Figure\,\ref{MSa}(c), and thus their linear extrapolation to zero
temperature also yields $H_c^{\| b}\simeq 13.9$~T. For $H>14.2$~T,
the $\alpha(T)$ curves show minima, whose positions shift to
higher $T$ with increasing field; see Figure\,\ref{TADTT}(b). In
Figure\,\ref{TADTT}(c), we compare the three curves closest to the
QCP on double-logarithmic scales. Moreover, a power-law fit of the
13.9~T curve is shown, which yields $\alpha(T,13.9~{\rm T})\propto
T^{-0.51}$, in almost perfect agreement with the expected
$1/\sqrt{T}$ divergence.

From the above analysis, one may expect the occurrence of sign
changes (minima) in $\alpha(T,H)$ at temperatures below our
minimum $T$ for the curves measured in fields between $H_c^{\| b}$ and
13.7~T (14.2~T). Because of the AFM ordering, however, this
behaviour will be definitely cut off at $T_\mathrm{N} \simeq
100\,\mathrm{mK}$, as it has been also observed in \hp
~\cite{RueggPRL08,ThielemannPRB09,Klanjsek08}. In any real
material, the question how well defined the QCP is, will depend on
such imperfections and also on the sample quality. As discussed in
Ref.~\cite{Anfuso08}, however, due to a magnetic correction of the
elastic moduli the quantum critical behavior may be cut off even
in a 'perfect' sample at low enough temperature by a first-order
phase transition.

%\section{Conclusion}

In conclusion, we have measured the low-temperature
magnetostriction and the thermal expansion along the spin-1/2
chain direction of Cu(C$_4$H$_4$N$_2$)(NO$_3$)$_2$ up to magnetic
fields strong enough to induce the quantum phase transition from
the Luttinger liquid phase to the fully spin-polarized state.
Using a Gr{\"u}neisen relation, our zero-field expansion data are well
described by the spin-1/2 Heisenberg chain hamiltonian. With
increasing field the experimental data follow the behaviour
expected on approaching a quantum critical point. In particular,
we observe the expected sign change and the $1/\sqrt{T}$
divergence of the thermal expansion.

\ack We thank A.\ V.\ Sologubenko and M.\ Garst for many fruitful
discussions and the Deutsche Forschungsgemeinschaft for the
financial support through SFB 608.

\end{document}